\documentclass[prd, amsfonts, twocolumn, nofootinbib, showpacs]{revtex4}
\usepackage{graphicx}
\usepackage{epsfig}
\usepackage{color}
\usepackage{amsmath}
\usepackage{amssymb}
\newcommand{\be}{\begin{equation}}
\newcommand{\ee}{\end{equation}}
\newcommand{\bea}{\begin{eqnarray}}
\newcommand{\eea}{\end{eqnarray}}

\newcommand{\gapp}{\mathrel{\raise.3ex\hbox{$>$}\mkern-14mu
\lower0.6ex\hbox{$\sim$}}}
\newcommand{\lapp}{\mathrel{\raise.3ex\hbox{$<$}\mkern-14mu
\lower0.6ex\hbox{$\sim$}}}

\def\bbox{{\,\lower0.9pt\vbox{\hrule \hbox{\vrule height 0.2 cm\hskip 0.2 cm \vrule  height 0.2 cm}\hrule}\,}}

\begin{document}
\title{A sensitive search for wormholes}
\author{John H. Simonetti$^1$, Michael J. Kavic$^2$,  
 Djordje Minic$^1$, Dejan Stojkovic$^3$, and De-Chang Dai$^{4,5}$}
\affiliation{$^1$ Department of Physics, Virginia Tech, Blacksburg, VA 24061, U.S.A.}
\affiliation{$^2$ Department of Chemistry and Physics, SUNY Old Westbury, Old Westbury, NY, U.S.A.}
\affiliation{ $^3$ HEPCOS, Department of Physics, SUNY at Buffalo, Buffalo, NY 14260-1500, U.S.A.}
\affiliation{$^4$ Center for Gravity and Cosmology, School of Physics Science and Technology, Yangzhou University, 180 Siwangting Road, Yangzhou City, Jiangsu Province, P.R. China 225002 }
\affiliation{ $^5$ CERCA/Department of Physics/ISO, Case Western Reserve University, Cleveland OH 44106-7079, U.S.A.}

\begin{abstract}
The realm of strong classical gravity and perhaps even quantum gravity are waiting to be explored.
In this letter we consider the recently detected triple system composed of two stars and a non-accreting black hole. Using published observations of this system we conduct the most sensitive test to date for whether the black hole is actually a wormhole by looking for orbital perturbations due to an object on the other side of the wormhole. The mass limit obtained on the perturber is $\sim4$ orders of magnitude better than for observations of S2 orbiting the supermassive black hole at Sgr A*. We also consider how observations of a pulsar could test for whether the black hole in a pulsar-black hole binary is a wormhole. A pulsar in a similar orbit to S2 would be $\sim10$ orders of magnitude more sensitive than observations of S2. For a nominal pulsar-black hole binary of stellar masses, with orbital size similar to that of the Hulse-Taylor binary pulsar, one year of observations could set a mass limit on a perturber that is $\sim6$ orders of magnitude better than observations of a pulsar around Sgr~A*. A range of limits between the pulsar-Sgr~A* and Hulse-Taylor cases could be obtained for a possible population of pulsar-black hole binaries that may exist near the galactic center. 
\end{abstract}

\maketitle

{\it Introduction:~} Might some black holes be wormholes? Black holes resulting from stellar evolution are not expected to be wormholes \cite{Misner:1974qy}. However, it has been argued that supermassive black holes may have a primordial formation history \cite{Duechting:2004dk}. Furthermore, even some stellar mass black holes in binary systems may be primordial \cite{Raidal:2018bbj}. It has been argued that primordial wormhole formation is possible and may be linked to primordial black hole formation \cite{Deng:2016vzb}. Recently it has even been claimed that a ninth planet (aside from Pluto) in the solar system might be primordial in nature \cite{Scholtz:2019csj}.

Can observations be used to test if specific black holes are wormholes? We explore a proposal, first discussed by \cite{Dai:2019mse}, to look for the effect of a perturbing object, orbiting on the other side of the wormhole, on the orbit of a star on our side (for other methods see e.g., \cite{Wang:2020emr,Dent:2020nfa,Liu:2020qia,Khodadi:2020jij,DeFalco:2020afv,Tangphati:2020mir,Jusufi:2020rpw,Godani:2020jpt,Tripathi:2019trz,Dokuchaev:2019jqq,Paul:2019trt}). Can we reasonably expect perturbers to orbit on the other side of a wormhole? It is well known that most stars are members of binaries or triple systems, etc. Thus, it is more likely that a stellar-mass black hole is a member of a multi-component system; an orbiting perturber on the other side of the wormhole is a reasonable scenario.

We consider two recently detected triple systems composed of two stars and a non-accreting black hole which can effectively be treated as a binary because of the distance of one of the stars from the black hole \cite{LB1,HD6819}. Observations of these systems would allow for the most sensitive search for a wormhole thus far conducted. We will also consider potential observations of pulsar-black hole (PSR-BH) binary systems, which can provide even more sensitive searches for a wormhole. Importantly, the existence of black hole-neutron star (BH-NS) systems has been confirmed by LIGO, at least in the case of preliminary candidates \cite{NSob}. Furthermore, a population of neutron star-black hole binaries is suggested to be present near the galactic center \cite{loeb}.  

The fascinating study of wormholes goes back to
Einstein and Rosen (ER) in 1935 \cite{ER}. This work was then explored in the 1950s and 1960s by John Wheeler \cite{geons}
and collaborators who have emphasized the importance of wormholes (and topology change) in quantum gravity \cite{wheeler}. 
In the 1980s Baum \cite{Baum:1984mc}, Hawking \cite{Hawking:1984hk} and Coleman \cite{Coleman:1988tj} focused on the role of topology change in Euclidean quantum gravity (see \cite{gibbons} for a review),
and they speculated that this process is crucial for the possible fix of fundamental constants in nature, and in particular,
the cosmological constant (see also \cite{Dai:2018vrw}). In a different research direction, but around the same time, 
Kip Thorne and collaborators realized that it was possible to construct
``traversable'' wormhole solutions \cite{Morris:1988cz, Morris:1988tu}. (For an illuminating review of this work
consult \cite{visser}.) More recently there has been a lot
of activity on the subject of wormholes and quantum entanglement
since the ER=EPR proposal \cite{Maldacena:2013xja} (see also, \cite{holland}, \cite{Dai:2020ffw}).

The first wormhole solution was originally constructed by Einstein and Rosen \cite{ER}. If we start from the static black hole metric in the Schwarzschild form
\begin{equation}
ds^2=-(1-\frac{2M}{r})dt^2+\frac{dr^2}{1-\frac{2M}{r}}+r^2d\Omega
\end{equation}
and apply a simple coordinate transformation, $u^2=r-2M$, we find
\begin{equation}
ds^2=-\frac{u^2}{u^2+2M}dt^2+4(u^2+2M)du^2+(u^2+2M)^2d\Omega.
\end{equation}
This metric contains two asymptotically flat spacetimes, $u>0$ and $u<0$, which are connected at $u=0$. This wormhole is non-traversable since it contains an event horizon, but it is not difficult to construct a traversable one, as we will do in the next section.

Where could such wormhole candidates come from?
One obvious source is the quantum gravity phase of the very early
universe. Even though such configurations would be exponentially
suppressed, inflation might make them macroscopic and thus
potentially observable. Their number has to be very small, so that
observed structure formation is not affected. Thus, observing such remnant
wormholes would be very challenging, but in principle feasible,
as explained in this letter.

In this paper we discuss a new constraint based on observations of a stellar orbit in a recently discovered triple system with a non-accreting BH, and even more stringent constraints that can be set using a pulsar-BH binary. Finally, we discuss observational prospects.

{\it Observable effects of a wormhole:~} It is a fascinating possibility that such a wormhole solution can be actually observed.
One approach has been recently addressed in \cite{Dai:2019mse}. We begin by summarizing this result.
We consider a simple wormhole model which can be studied analytically. 
A standard Schwarzschild space-time metric with the gravitational radius $r_g=2G/M$ is given as
\begin{equation}
ds^2=-(1-\frac{r_g}{r})dt^2+\frac{dr^2}{1-\frac{r_g}{r}}+r^2d\Omega .
\end{equation}
We cut this space-time at the radius $R$ which is slightly bigger than the gravitational radius, i.e. $R \ge r_g$. We take another identical space-time and paste them together.
Our global construct is thus  two copies of the Schwarzschild space-time connected through a mouth of radius $R$. This setup represents a short throat wormhole, which is traversable since $R \ge r_g$. Some exotic matter with negative energy density is needed to keep the wormhole open, however, in the short throat approximation that we use, we assume that the effects of this exotic matter are subdominant.  

We can now study perturbations in this background.
We label the radial coordinate in ``our space" where the objects we observe are located with $r_2$,  and the radial coordinate in the ``other space"  where the source of perturbations is located with $r_1$. Outside of the mouth, i.e., for $r_1>R$ and $r_2>R$, the space-time is Schwarzschild on both sides. These two copies of the Schwarzschild space-time are connected at $r_1=r_2=R$. 

The monopole metric perturbations in the Schwarzschild background can be written as \cite{1966PhRv..146..938P,Zerilli:1971wd,Garat:1999vr,Detweiler:2003ci,Barack:2005nr,Chen:2016plo}

\begin{eqnarray}
\label{perturbation1}
h_{tt}&=&\frac{2\mu}{r}\Theta(r-A)+\frac{2\mu}{A}\Theta(A-r)\\
\label{perturbation2}
h_{rr}&=&\frac{2\mu r }{(r-r_g)^2}\Theta(r-A)
\end{eqnarray}
where $\mu$ is the  mass of an object that perturbs the metric, while $A$ is its location. $\Theta(x)$ is the standard Heaviside function. This will be a good starting point since we are working in the short-throat wormhole approximation. 

In the wormhole spacetime, we  write the perturbations in the ``other space" as
\begin{eqnarray}
h^{oth}_{tt}(r_1)&=&h_{tt}(r_1)+\frac{2a_{tt}}{r_1}\\
h^{oth}_{rr}(r_1)&=&h_{rr}(r_1)+\frac{2a_{rr} r_1 }{(r_1-r_g)^2} .
\end{eqnarray}
From Eqs.~\eqref{perturbation1} and \eqref{perturbation2}, we see that $a_{tt}$ and $a_{rr}$ are not completely independent. They are both equal to an effective mass in the ``other space," $a_{tt}=a_{rr}=\mu^{oth}$. The perturbations in ``our space" are
\begin{eqnarray}
h^{our}_{tt}(r_2)&=&\frac{2b_{tt}}{r_2}\\
h^{our}_{rr}(r_2)&=&\frac{2b_{rr} r_2 }{(r_2-r_g)^2} .
\end{eqnarray}
Again $b_{tt}$ and $b_{rr}$ are not completely independent, and they are both equal to an effective mass in our space, $b_{tt}=b_{rr}=\mu^{our}$.
Note that the terms $a_{tt}$, $a_{rr}$, $b_{tt}$ and $b_{rr}$ are added to account for the presence of the wormhole. 
We will find their concrete forms by matching the perturbations at the the wormhole mouth.
We require that $h_{tt}$ is continuous at $r=R$, so the time variable is the same inside and outside of the shell $r=R$. 
Because the source of perturbations is not located at the mouth, we also require that the derivative of $h_{tt}$ is continuous at $r=R$.
 Thus, from the continuity conditions $h^{our}_{tt}(R)=h^{oth}_{tt}(R)$ and $\partial_{r_2}h^{our}_{tt}\vert_{r_2=R}=\partial_{r_1}h^{our}_{tt}\vert_{r_1=R}$ we find
\begin{eqnarray}
b_{tt}&=&-a_{tt}=\mu \frac{R}{A}\\
b_{rr}&=&-a_{rr}=\mu \frac{R }{A} .
\end{eqnarray}

Since $b_{tt}$ is nonzero, an observer on our side can feel an additional acceleration due to perturbations sourced on the other side. If the observer is far away from the wormhole, this additional acceleration is
\begin{equation}
\label{acceleration}
a\approx-\mu \frac{R}{A}\frac{1}{r_2^2} .
\end{equation}

If all we had was just a monopole contribution, it would be very difficult to extract an observable effect, since this additional acceleration would just simply add to the acceleration due to the central object. We therefore consider an elliptic orbit of a perturber, i.e., an object orbiting on the other side of the wormhole with the periapsis radius $r_p$ and apoapsis radius $r_a$. An elliptic orbit cannot be represented with only one monopole, and can be viewed instead as a sequence of monopoles. We estimate the magnitude of the acceleration variation by using two monopoles, one for a perigee, $r_p$, and another for an apogee, $r_a$, as 
\begin{equation}
\Delta a =\mu R \left(\frac{1}{r_p}-\frac{1}{r_a}\right)\frac{1}{r_2^2} .
\end{equation}
If the orbit of an object on the other side of the wormhole's is elongated so that $r_a\gg r_p$, then we can approximate the magnitude as
\begin{equation}\label{da}
\Delta a =\mu \frac{R}{r_p}\frac{1}{r_2^2} .
\end{equation}

Note that what we calculate in Eq.~(\ref{da}) is the magnitude of acceleration variation of an object in our space due to an elliptic orbit of a perturber on the other side perturbing the metric. These variations come on top of the constant acceleration that
comes from the central object. With good enough precision, we should be able to detect or exclude this variable anomalous acceleration. These variations could be produced by some other dim sources on our side. Then, more careful modeling would be required to distinguish between different options.

It is important to note that our wormhole has Schwarzschild geometry outside of the mouth, while the horizon is not present at all, since we cut the Schwarzschild geometry at $R>r_g$.
Thus, such wormholes can be  harbored both by black hole candidates (either stellar mass of super-massive ones) and/or other compact objects less massive than black holes. In particular, a neutron star candidate might as well be a wormhole, as long as we do not see its surface.  

{\it Searching for wormholes:~} Dai and Stojkovic \cite{Dai:2019mse} considered observations of the star S2 in orbit around the supermassive BH at the center of our Galaxy, at Sgr~A*, to produce tentative limits on the perturber, if the BH is a wormhole. 

The most direct way to observe the effect of the anomalous acceleration shown in Eq.~(\ref{da}) is to look for deviations of the star's orbit from the expected Keplerian result. We expect the most sensitive method is to test for secular orbital effects (e.g., a secular change in the orbital period, or the advance of the periapsis through precession). We consider a secular change in orbital period. 

To estimate the change in the orbital period caused by $\Delta a$ given in Eq.~(\ref{da}) we assume, for simplicity, that the additional acceleration occurs once every orbital period $T$ of the perturber (i.e., when it is near its periapsis). We will consider systems where the duration of the additional acceleration is $t_p \ll P_b$, where $P_b$ is the binary orbital period of the star on our side of the wormhole, and so we treat the effect of the perturber as impulsive.

We estimate the resulting change in the star's orbital energy (per unit mass), caused by \textit{one} such impulse, as
\begin{equation}
    \delta E\sim
    \Delta a\, \hat{a} \cdot \vec{v}_{star}\, t_p \sim
    \pm \Delta a\, \sqrt{ \frac{GM_{BH}}{a_b} }\, t_p,
\end{equation}
where $t_p$ is the time the perturber spends near periapsis, $\hat{a}$ is the unit vector in the direction of the anomalous acceleration, $\vec{v}_{star}$ is the star's velocity vector, and $a_b$ is the semi-major axis of the star's orbit. This estimate assumes such an impulse occurs at a random moment during the star's orbit, not when the star is at its periapsis or apoapsis. If the impulse occurs as the star is approaching periapsis, its orbital energy and period increase. If it occurs as the star recedes from periapsis the energy and period decrease. 

To estimate $t_p$ we note $T=t_p+t_a\sim t_a$ where $t_a$ is the time the perturber spends away from periapsis (i.e., mostly at apoapsis for $r_a\gg r_p$). So, where $v_p$ and $v_a$ are the periapsis and apoapsis speeds of the perturber, respectively, we have
\begin{equation}
    t_p \sim t_p \frac{T}{t_a}
    \sim \frac{r_p}{v_p} \frac{v_a}{r_a} T
    \sim \left(\frac{r_p}{r_a}\right)^2 T \sim f^2 T
\end{equation}
where $f=r_p/r_a$, and we used $v_a r_a = v_p r_p$ by conservation of angular momentum.

A change in the star's orbital energy $\delta E$ produces a change in the semi-major axis $\delta a_b$ (through $E=-GM_{BH}/2a_b$) and therefore a change in orbital period, via Kepler's third law. The result for one impulse is a magnitude of change in the period of
\begin{equation}
    \delta P_b \sim 6\pi \frac{M}{M_{BH}} \frac{r_g}{r_p} f^2 T,
\end{equation}
where $M$ is the mass of the perturber, we take $R\sim r_g$ as in \cite{Dai:2019mse}, and we have used $r_2\sim a_b$.

We will consider only the case where $P_b \gg T$. Then many impulses increasing $P_b$ occur as the star approaches its periapsis, and an almost equal number of impulses decreasing $P_b$ occur as the star recedes from periapsis. The positive and negative changes will {\it almost} cancel for a full orbit, but not exactly so, since the star and perturber are not likely to reach their respective periapsis points at the same time. The net effect is an increase (or decrease) of the orbital period of the star by an amount roughly equal to the effect of one impulse. Then, since the effect on the star's orbit is small for the perturbers we are considering, the orbit does not change substantially and the imbalance persists for subsequent orbits. The net effect is a secular change in the star's orbital period of magnitude
\begin{equation}
    \Delta P_b \sim \delta P_b \frac{\Delta T}{P_b}
\end{equation}
where $\Delta T$ is the duration of the observing program. This effect is observable if $\Delta P_b$ is larger than the measurement precision on the period. Otherwise a limit on the mass of the perturber can be set as
\begin{equation}
    M\sim 
    \frac{M_{BH}}{6\pi} \frac{\sigma_{P_b}}{f^2T}
    \frac{r_p}{r_g} \frac{P_b}{\Delta T}
\end{equation}
where $\sigma_{P_b}$ is the measurement error in the star's orbital period. Since $T\propto r_p^{3/2}$, this limit is $M\propto r_p^{-1/2}$.

\begin{figure}
    \centering
    \includegraphics[width=85mm]{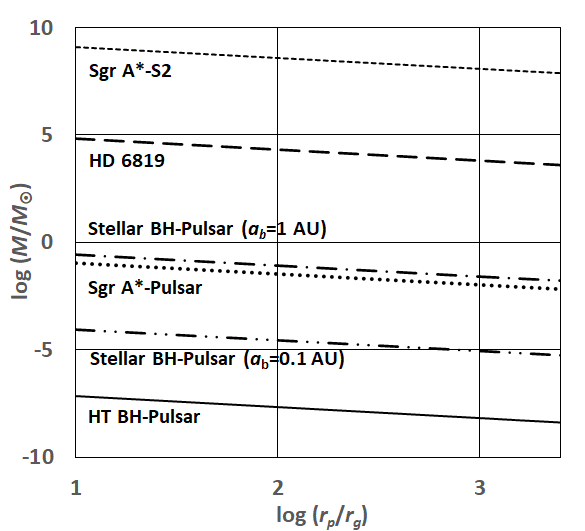}
    \caption{The mass limit on the perturber as a function of its periapsis distance from the wormhole (expressed in units of the gravitational radius of the BH/wormhole).
    The upper two lines are for existing observations of an ordinary star orbiting a supermassive BH or stellar-mass black hole.
    The other lines show limits that could be set for observations of a pulsar orbiting a supermassive BH or stellar-mass back hole.
    The HT BH-Pulsar case is a Hulse-Taylor-sized stellar-mass BH-Pulsar binary system.}
    \label{fig:limit}
\end{figure}

For numerical results we first consider the case of S2 orbiting the BH at Sgr~A*. From 20 years of observations of S2 (more than 1 orbit, necessary for any reasonable measurement of the orbital period) we have $P_b=15.92\pm0.04$~years \cite{boehle}. For all the cases we consider in this paper we take $f=0.1$. For $M_{BH}=4\times10^6M_\odot$ we obtain a mass limit for a perturber as a function of $r_p/r_g$ given by the upper-most line shown with short dashes in Fig.~\ref{fig:limit}.

A better limit can be set from existing observations of a star in orbit around a stellar-mass BH, instead of the supermassive BH at Sgr~A*, specifically the orbit of the B3III star around the recently discovered non-accreting, stellar-mass BH in the triple system HD~6819, noted above. Given the observed period of this inner binary (the third component is very distant) is $P_b=40.333\pm0.004$~days, and the black hole $M_{BH}=4.2 M_\odot$ \cite{HD6819}, the perturber mass limit for this case is $\sim$4 orders of magnitude lower than obtained from observations of S2, and is shown by the long-dash line in Fig.~\ref{fig:limit}, for $\Delta T=1$~year. This limit represents the most sensitive search for a wormhole to date. A similar constraint can be set using observations of the B star LB-1 in another triple containing a non-accreting BH \cite{LB1}, but the mass limit for HD~6819 system is better by a factor of $\sim2$.

However, observations of a pulsar orbiting a black hole have the potential to set even better limits, given the greater observational precision obtainable. BH-pulsar binaries have been argued to provide remarkable tests of quantum gravity \cite{mike1,mike2,mike3,mike4,mike5,mike6,mike7} on top of their proven record in testing Einstein's general relativity in the case of the Hulse-Taylor BH-pulsar binary PSR~B1913$+$16 \cite{weisberg}. The precision on measured orbital parameters for pulsars is determined by the precision on pulse ``times of arrival'' (TOA) measurements, which is typically $\sigma_{TOA}\sim1\mu$s \cite{condon}. A pulsar TOA measurement is obtained from $\sim10$ minutes of data at each observing epoch (during which a folding and pulse-shape averaging process is applied). The result is one TOA for that epoch; successive TOAs are fed into a software package such as TEMPO (www.pulsar.princeton.edu/tempo) which models the pulsar's behavior yielding various parameters describing the pulsar, including $P_b$, with formal uncertainties. To estimate $\sigma_{P_b}$ for a pulsar in a binary system we used discussions of modeling using epoch-folding for periodic phenomena \cite{larsson}\cite{kovacs}. The expression
\begin{equation} \label{pulsarerror}
    \sigma_{P_b} \approx 
    \frac{\sqrt{6}}{\pi}
    \frac{\sigma_{TOA}\, P_b}
    { (a_b\,\sin i/c)
    \sqrt{N_{TOA}} 
    }
    \frac{P_b}{\Delta T},
\end{equation}
where $i$ is the inclination angle of the orbit and $N_{TOA}$ is the number of TOA measurements,
is adapted from \cite{larsson} and gives a reasonable approximation to the TEMPO result (e.g., in the case of the Hulse-Taylor pulsar). Observations of at least a full orbit are needed to obtain this precision on the period. 

For a pulsar in an orbit around Sgr A* similar to that of S2, using Eq.~(\ref{pulsarerror}) for the uncertainty in the measured orbital period for the pulsar with $a_p\sin i\sim 1000$~AU and $\Delta T=20$~years, we obtain a mass limit for the perturber that is $\sim10$ orders of magnitude lower than for observations of S2. The result is the dotted line in Fig.~\ref{fig:limit}.

Still better results could be obtained for pulsars in close orbits around stellar-mass black holes. Consider the ``nominal'' case of a pulsar in orbit around a $10M_\odot$ BH where $r_p$ and $r_2$ are equal to the semi-major axis for the Hulse-Taylor pulsar, $a_b\approx2\times10^9$~m. For observations over $\Delta T=1$~year, using the orbital period uncertainty given by Eq.~(\ref{pulsarerror}) we obtain a limit on the perturber mass that is $\sim6$ orders of magnitude better than for a pulsar orbiting Sgr~A*, shown as the solid line in Fig.~\ref{fig:limit}.

Lastly, we consider a population of BH-pulsar binaries that may be present in the galactic center \cite{loeb}. The semi-major axes of these binaries would range from $a_p\sim0.1$~AU to $a_p\sim1$~AU, with eccentricities $\sim0.8$. The perturber mass limits attainable for these systems are near to, or below the limit for a pulsar orbiting Sgr~A*, but not as low as the nominal Hulse-Taylor-sized pulsar-BH binary. These results are also shown in Fig.~\ref{fig:limit}.

{\it Observational Prospects and Outlook:~}
The best prospects for identifying stable BH-NS systems stem from either gravitational wave detection with a follow-up search for pulsar emission, or the direct detection of pulsars in a binary system followed by determination of the nature of the binary partner. 
LISA is designed to detect stable binary systems including BH-NS systems \cite{AmaroSeoane:2012km}. The SKA is designed to be able to detect all the pulsars in our galaxy including near the galactic center where BH-pulsar systems may be more common \cite{Lipunov:2005sv}.
In future work we plan to use numerical simulations to further explore the perturber limits that can be obtained. We will also explore connections with the recent research on quantum gravity/string theory \cite{Freidel:2013zga} with intrinsic non-locality that could be probed as outlined in this letter.

{\it Acknowledgments:} D-C.\ Dai was supported by the National Science Foundation of China (Grant No.\ 11433001 and 11775140), National Basic Research Program of China (973 Program 2015CB857001) and  the Program of Shanghai Academic/Technology Research Leader under Grant No.\ 16XD1401600. D.M.\ is supported by the Julian Schwinger Foundation and the Department of Energy (under grant DE-SC0020262). D.S.\ was partially supported by the US National Science Foundation Grant No.\ PHY-1820738.




\begin{thebibliography}{99}

\bibitem{Misner:1974qy}
C.~W.~Misner, K.~S.~Thorne and J.~A.~Wheeler,
 {\sl Gravitation}, Princeton Univ. Press, Princeton, 2017.

\bibitem{Duechting:2004dk}
N.~Duechting,
Phys. Rev. D \textbf{70}, 064015 (2004)
doi:10.1103/PhysRevD.70.064015
[arXiv:astro-ph/0406260 [astro-ph]].


\bibitem{Raidal:2018bbj}
M.~Raidal, C.~Spethmann, V.~Vaskonen and H.~Veermäe,
JCAP \textbf{02}, 018 (2019)
doi:10.1088/1475-7516/2019/02/018
[arXiv:1812.01930 [astro-ph.CO]].

\bibitem{Deng:2016vzb}
H.~Deng, J.~Garriga and A.~Vilenkin,
JCAP \textbf{04}, 050 (2017)
doi:10.1088/1475-7516/2017/04/050
[arXiv:1612.03753 [gr-qc]].

\bibitem{Scholtz:2019csj}
J.~Scholtz and J.~Unwin,
[arXiv:1909.11090 [hep-ph]].

\bibitem{Dai:2019mse} 
  D.~C.~Dai and D.~Stojkovic,
  Phys.\ Rev.\ D {\bf 100}, no. 8, 083513 (2019);
D.~C.~Dai and D.~Stojkovic,
  ``Response to the Comment on "Observing a wormhole",''
  arXiv:1912.07793 [gr-qc].




\bibitem{Dent:2020nfa}
J.~B.~Dent, W.~E.~Gabella, K.~Holley-Bockelmann and T.~W.~Kephart,
[arXiv:2007.09135 [gr-qc]].

\bibitem{Wang:2020emr}
X.~Wang, P.~C.~Li, C.~Y.~Zhang and M.~Guo,
[arXiv:2007.03327 [gr-qc]].


\bibitem{Liu:2020qia}
H.~Liu, P.~Liu, Y.~Liu, B.~Wang and J.~P.~Wu,
[arXiv:2007.09078 [gr-qc]].



\bibitem{Khodadi:2020jij}
M.~Khodadi, A.~Allahyari, S.~Vagnozzi and D.~F.~Mota,
[arXiv:2005.05992 [gr-qc]].

\bibitem{DeFalco:2020afv}
V.~De Falco, E.~Battista, S.~Capozziello and M.~De Laurentis,
Phys. Rev. D \textbf{101}, no.10, 104037 (2020)
doi:10.1103/PhysRevD.101.104037
[arXiv:2004.14849 [gr-qc]].

\bibitem{Tangphati:2020mir}
T.~Tangphati, A.~Chatrabhuti, D.~Samart and P.~Channuie,
[arXiv:2003.01544 [gr-qc]].



\bibitem{Jusufi:2020rpw}
K.~Jusufi, P.~Channuie and M.~Jamil,
Eur. Phys. J. C \textbf{80}, no.2, 127 (2020)
doi:10.1140/epjc/s10052-020-7690-7
[arXiv:2002.01341 [gr-qc]].

\bibitem{Godani:2020jpt}
N.~Godani, S.~Debata, S.~K.~Biswal and G.~C.~Samanta,
Eur. Phys. J. C \textbf{80}, no.1, 40 (2020)
doi:10.1140/epjc/s10052-019-7596-4



\bibitem{Tripathi:2019trz}
A.~Tripathi, B.~Zhou, A.~B.~Abdikamalov, D.~Ayzenberg and C.~Bambi,
Phys. Rev. D \textbf{101}, no.6, 064030 (2020)
doi:10.1103/PhysRevD.101.064030
[arXiv:1912.03868 [gr-qc]].

\bibitem{Dokuchaev:2019jqq}
V.~I.~Dokuchaev and N.~O.~Nazarova,
[arXiv:1911.07695 [gr-qc]].

\bibitem{Paul:2019trt}
S.~Paul, R.~Shaikh, P.~Banerjee and T.~Sarkar,
JCAP \textbf{03}, 055 (2020)
doi:10.1088/1475-7516/2020/03/055
[arXiv:1911.05525 [gr-qc]].










\bibitem{HD6819} Th. Rivinius, D. Baade, P. Hadrava, M. Heida and R. Klement, Astron. \& Astrophy., {\bf 639}, L3 (2020).

\bibitem{LB1} J. Liu, H. Zhang, H. Howard et al., Nature, {\bf 575}, 618 (2019).



\bibitem{NSob}
{The LIGO Scientific Collaboration and the Virgo Collaboration}.
  2019{\natexlab{b}}, GRB Coordinates Network, 25333, 1


\bibitem{loeb} C. Faucher-Gigu\`ere and A. Loeb, MNRAS, {\bf 415}, 3951 (2011).






\bibitem{ER} A. Einstein and N. Rosen, Phys. Rev. {\bf 48}, 73 (1935).

\bibitem{geons} J. A. Wheeler, 
Phys.\ Rev.\ {\bf 97}, 511 (1955).

\bibitem{wheeler} J. A. Wheeler, {\it Geometrodynamics}, Academic, New York, 1962.

\bibitem{Baum:1984mc}
  E.~Baum,
  Phys.\ Lett.\  {\bf 133B}, 185 (1983).

\bibitem{Hawking:1984hk}
  S.~W.~Hawking,
  Phys.\ Lett.\  {\bf 134B}, 403 (1984).

\bibitem{Coleman:1988tj}
  S.~R.~Coleman,
  Nucl.\ Phys.\ B {\bf 310}, 643 (1988).


\bibitem{gibbons}
G. W. Gibbons and S. W. Hawking (editors), {\it Euclidean Quantum Gravity}, World Scientific, 1993.

\bibitem{Dai:2018vrw} 
  D.~C.~Dai, D.~Minic and D.~Stojkovic,
  Phys.\ Rev.\ D {\bf 98}, no. 12, 124026 (2018)


\bibitem{Morris:1988cz}
  M.~S.~Morris and K.~S.~Thorne,
  Am.\ J.\ Phys.\  {\bf 56}, 395 (1988).

\bibitem{Morris:1988tu}
  M.~S.~Morris, K.~S.~Thorne and U.~Yurtsever,
  Phys.\ Rev.\ Lett.\  {\bf 61}, 1446 (1988).


\bibitem{visser} M. Visser, {\it Lorentzian Wormholes: From Einstein to Hawking}, AIP Press, New York, 1995


\bibitem{Maldacena:2013xja}
  J.~Maldacena and L.~Susskind,
  Fortsch.\ Phys.\  {\bf 61}, 781 (2013)


\bibitem{holland}
In P. R. Holland's book,  {\it Quantum Theory of Motion}, (Cambridge 1995), a connection between $ER$ and $EPR$ has been
suggested in the context of the de-Broglie-Bohm interpretation of quantum theory.


\bibitem{Dai:2020ffw} 
  D.~C.~Dai, D.~Minic, D.~Stojkovic and C.~Fu,
  ``Testing ER=EPR,''
  arXiv:2002.08178 [hep-th].










 \bibitem[Peters(1966)]{1966PhRv..146..938P} Peters, P.~C.\ 1966, Physical Review, 146, 938

\bibitem{Zerilli:1971wd}
  F.~J.~Zerilli,
  Phys.\ Rev.\ D {\bf 2}, 2141 (1970).
  doi:10.1103/PhysRevD.2.2141

\bibitem{Garat:1999vr}
  A.~Garat and R.~H.~Price,
  Phys.\ Rev.\ D {\bf 61}, 044006 (2000)
  doi:10.1103/PhysRevD.61.044006
  [gr-qc/9909005].

\bibitem{Detweiler:2003ci}
  S.~L.~Detweiler and E.~Poisson,
  Phys.\ Rev.\ D {\bf 69}, 084019 (2004)
  doi:10.1103/PhysRevD.69.084019
  [gr-qc/0312010].


\bibitem{Barack:2005nr}
  L.~Barack and C.~O.~Lousto,
  Phys.\ Rev.\ D {\bf 72}, 104026 (2005)
  doi:10.1103/PhysRevD.72.104026
  [gr-qc/0510019].

\bibitem{Chen:2016plo}
  J.~E.~Thompson, B.~F.~Whiting and H.~Chen,
  Class.\ Quant.\ Grav.\  {\bf 34}, no. 17, 174001 (2017)
  doi:10.1088/1361-6382/aa7f5b
  [arXiv:1611.06214 [gr-qc]].

\bibitem{boehle}
A. Boehle, et al., Ap.J., 830, 17, 2016.


\bibitem{mike1}
M.~Kavic, J.~H.~Simonetti, S.~E.~Cutchin, S.~W.~Ellingson and C.~D.~Patterson,
  JCAP {\bf 0811}, 017 (2008).
  
\bibitem{mike2}
M.~Kavic, D.~Minic and J.~Simonetti,
  Int.\ J.\ Mod.\ Phys.\ D {\bf 17}, 2495 (2009).
  
\bibitem{mike3}
J.~H.~Simonetti, M.~Kavic, D.~Minic, U.~Surani and V.~Vijayan,
  Astrophys.\ J.\  {\bf 737}, L28 (2011).
  
\bibitem{mike4}
J.~Estes, M.~Kavic, M.~Lippert and J.~H.~Simonetti,
  Int.\ J.\ Mod.\ Phys.\ D {\bf 26}, no. 12, 1743004 (2017).
  
\bibitem{mike5}
S.~L.~Liebling, M.~Lippert and M.~Kavic,
  JHEP {\bf 1803}, 176 (2018).
  
\bibitem{mike6}
M.~J.~Kavic, D.~Minic and J.~Simonetti,
  Int.\ J.\ Mod.\ Phys.\ D {\bf 27}, no. 14, 1847007 (2018).
  
\bibitem{mike7}
 M.~Kavic, S.~L.~Liebling, M.~Lippert and J.~H.~Simonetti,
  arXiv:1910.06977 [astro-ph.HE].

\bibitem{weisberg} J.M. Weisberg and Y. Huang, Ap. J., {\bf 829}, 55 (2016).

\bibitem{condon}
J.J. Condon and S.M. Ransom, {\sl Essential Radio Astronomy}, Princeton Univ. Press, Princeton, 2016.

\bibitem{larsson}
S. Larsson, Astr. Ap. Suppl., 117, 197, 1996.

\bibitem{kovacs}
G. Kovacs, Ap. Sp. Sci., 78, 175, 1981.

\bibitem{S2precession} R.~Abuter et al., A\&A, {\bf 636}, L5 (2020).

\bibitem{AmaroSeoane:2012km}
P.~Amaro-Seoane, et al.,
GW Notes \textbf{6}, 4-110 (2013)
[arXiv:1201.3621 [astro-ph.CO]].


\bibitem{Lipunov:2005sv}
V.~Lipunov, A.~Bogomazov and M.~Abubekerov,
Mon. Not. Roy. Astron. Soc. \textbf{359}, 1517-1523 (2005)
doi:10.1111/j.1365-2966.2005.08997.x
[arXiv:astro-ph/0503341 [astro-ph]].


\bibitem{Freidel:2013zga}
L.~Freidel, R.~G.~Leigh and D.~Minic,
  Phys.\ Lett.\ B {\bf 730}, 302 (2014)
  Int.\ J.\ Mod.\ Phys.\ D {\bf 23}, no. 12, 1442006 (2014)
  JHEP {\bf 1506}, 006 (2015)
  Int.\ J.\ Mod.\ Phys.\ D {\bf 24}, no. 12, 1544028 (2015).
  Phys.\ Rev.\ D {\bf 94}, no. 10, 104052 (2016)
  J.\ Phys.\ Conf.\ Ser.\  {\bf 804}, no. 1, 012032 (2017).
  JHEP {\bf 1709}, 060 (2017)
  Phys.\ Rev.\ D {\bf 96}, no. 6, 066003 (2017)
  Int.\ J.\ Mod.\ Phys.\ A {\bf 34}, no. 28, 1941004 (2019).
  L.~Freidel, J.~Kowalski-Glikman, R.~G.~Leigh and D.~Minic,
  Phys.\ Rev.\ D {\bf 99}, no. 6, 066011 (2019)

\end{thebibliography}
\end{document}